\documentclass[runningheads]{llncs} 

\usepackage{graphicx}
\usepackage[usenames,dvipsnames,svgnames,table]{xcolor}
\usepackage{mathtools}
\usepackage{amssymb,amsmath}
\usepackage{ifthen}
\usepackage{color}
\usepackage{xspace}
\usepackage{listings}
\usepackage{float}
\usepackage{makecell}
\usepackage{wrapfig}
\usepackage{lipsum}
\usepackage{array}
\usepackage[utf8]{inputenc}
\usepackage{textcomp}
\usepackage{enumitem}
\usepackage{booktabs}
\usepackage[final]{pdfpages}
\usepackage[colorlinks=true,linkcolor=blue,urlcolor=blue,citecolor=blue]{hyperref} 
\usepackage[section]{placeins}
\usepackage{tikz}
\usetikzlibrary{decorations.pathmorphing, patterns}
\usepackage[scriptsize]{subfigure}
\usepackage[small,bf]{caption}
\usepackage{pifont}

\usepackage{wrapfig}
\usepackage{caption}
\usepackage[abbreviations]{foreign}
\usepackage{pslatex}

\usepackage{tabularx}
\newcolumntype{L}[1]{>{\raggedright\arraybackslash}p{#1}}
\newcolumntype{C}[1]{>{\centering\arraybackslash}p{#1}}
\newcolumntype{R}[1]{>{\raggedleft\arraybackslash}p{#1}}

\usepackage{hyphenat}
\newboolean{showcomments}
\setboolean{showcomments}{false}
\ifthenelse{\boolean{showcomments}}
{ \newcommand{\mynote}[3]{
    \fbox{\bfseries\sffamily\scriptsize#1}
    {\small$\blacktriangleright$\textsf{\emph{\color{#3}{#2}}}$\blacktriangleleft$}}}
{ \newcommand{\mynote}[3]{}}
\newcommand{\vs}[1]{\mynote{Valerio}{#1}{blue}}
\newcommand{\ob}[1]{\mynote{Oscar}{#1}{red}}

\newcommand{\tz}{\textsc{TrustZone}\xspace}
\newcommand{\arm}{\textsc{Arm}\xspace}
\newcommand{\optee}{\textsc{Op-Tee}\xspace}
\newcommand{\sgx}{Intel\textregistered SGX\xspace}
\newcommand{\mqtt}{\textsc{Mqtt}\xspace}
\newcommand{\mosquitto}{\texttt{mosquitto}\xspace}
\newcommand{\mqttz}{\textsc{Mqt-Tz}\xspace}

\newcommand{\sys}{\textsc{KeVlar-Tz}\xspace}

\title{\sys: a Secure Cache for \arm \tz\\\normalsize(Practical Experience Report)}

\titlerunning{\sys: a Secure Cache for \arm \tz}

\author{Oscar Benedito\inst{1}, Ricard Delgado-Gonzalo\inst{1}\orcidID{0000-0002-7183-6257} \and Valerio Schiavoni\inst{2}\orcidID{0000-0003-1493-6603}}
\institute{
CSEM, Neuch\^atel, Switzerland,
\email{{{obo,rdg}@csem.ch}}
\and
University of Neuch\^atel, Switzerland,
\email{valerio.schiavoni@unine.ch}
}

\begin{document}

\maketitle
\begin{abstract}
Edge devices are increasingly in charge of storing privacy-sensitive data, in particular implantables, wearables, and nearables can potentially collect and process high-resolution vital signs 24/7.
Storing and performing computations over such data in a privacy-preserving fashion is of paramount importance.
We present \sys, an application-level trusted cache designed to leverage \arm\tz, a popular trusted execution environment available in consumer-grade devices.
To facilitate the integration with existing systems and IoT devices and protocols, \sys exposes a REST-based interface with connection endpoints inside the \tz enclave. 
Furthermore, it exploits the on-device secure persistent storage to guarantee durability of data across reboots.
We fully implemented \sys on top of the \optee framework, and experimentally evaluated its performance. 
Our results showcase performance trade-offs, for instance in terms of throughput and latency, for various workloads, and we believe our results can be useful for practitioners and in general developers of systems for \tz.
\sys is available as open-source at \url{https://github.com/mqttz/kevlar-tz/}.

\end{abstract}

\vspace{-4pt}
\section{Introduction} \label{sec:intro}
\vspace{-4pt}
Wearable and Internet-of-Things (IoT) devices are becoming increasingly pervasive in modern society.
It is predicted that by the year 2025 there will be more than 600 million wearable devices deployed and connected worldwide~\cite{global-wearable}, and according to Cisco up to 500 billion IoT devices by 2030~\cite{cisco-iot}.
These devices continuously produce data from a wide range of sensor types: inertial sensors (\eg accelerometers, gyroscopes)~\cite{Bennett2016}, biopotential (\eg electrocardiography)~\cite{Chaudhuri2009}, optical (\eg photopletysmography)~\cite{Tamura2014}, biochemical (\eg pH)~\cite{Coyle2014}, \etc. Combinations of such sensors allow for the monitoring of the health statuses of the users, ranging from the user's physical activity~\cite{DelgadoGonzalo2017} to the detection of cardiac abnormalities~\cite{Faraone2020}.
The nature of this data is intrinsically privacy-sensitive.
Applications and system designers must protect it from malicious attackers, including those with physical access, from accessing and possibly unveiling them.
Similarly, IoT devices are regularly used to monitor and record privacy-related data.
Examples include motion sensors (\eg, in the case of a smart-home deployment, revealing for instance the presence of humans indoors~\cite{lin2016iot}), power-consumption meters (\eg, potentially revealing the habits of a household), weather sensors (\eg, a key asset in farming used to decide on optimal irrigation levels~\cite{padalalu2017smart}), \etc.
The vast majority of such applications deal with the insertion and retrieval of data from/to a dedicated, and preferably persistent, memory area.
The mentioned operations are typically offered by key-value stores, \eg, software libraries or services that allow to \texttt{put} and \texttt{get} values associated with unique identifiers (\ie, the keys), for later retrieval, similar to a \emph{caching} mechanism.
Note that such libraries are vastly known in literature (\ie \cite{246158,han2011survey}, \etc), extensively studied~\cite{gokhale2010kvzone} and find usage in several and diverse application domains.
Noteworthy, the result of confidential computations (\eg, edge-based privacy-preserving machine-learning model training, just to name one) must also be stored and retrieved following the same access patterns.
Hence, the content of such memory area must be shielded.

The need for stringent data privacy guarantees, such as the mentioned shielding, usually comes at the cost of computational overhead.
This is the case of full homomorphic encryption (HE)~\cite{gentry2009fully}, a purely software-based approach to compute and operate over encrypted data.
However, recent work~\cite{gottel2018security} has shown how state-of-the-art HE implementations~\cite{halevi2013design} still result in orders of magnitude slowdown even for simple arithmetical operations, and major breakthroughs are yet to be seen for HE to become a viable solution.

The introduction and widespread adoption in the last few years of trusted execution environments (TEE) for consumer- and server-grade devices offers an opportunity to combine the need for privacy with the ones of viable performance.
TEEs provide a hardware-supported mechanism to maintain the privacy and integrity of data while allowing for efficient and transparent protection from malicious attackers or compromised operating systems.
Such protected areas are commonly referred to as \emph{enclaves}, and they represent the main programming abstraction supported by the large majority of available TEEs.
Notable examples include \sgx~\cite{costan2016intel}, AMD Secure Encrypted Virtualization (SEV)~\cite{kaplan2016amd} for server-grade as well as cloud-based deployments~\cite{gce-sev,sgx-aws} and \arm \tz~\cite{amacher2019performance,pinto2019demystifying} for more edge-centric scenarios, the focus of this work.

In this practical experience report paper, we present \sys, an efficient trusted cache application for \arm \tz with support for non-volatile secure storage.
\sys exposes an easy-to-use REST interface to facilitate the integration with existing systems, protocols, and third-party devices.
The network connection endpoints are established within the \tz enclave.
Finally, \sys is designed to exploit the secure storage implemented by some \tz-enabled systems, allowing for secure data durability.

The main \textbf{contributions} of this work are twofold.
First, we present the design and implementation of \sys, a secure cache for \arm \tz.
Second, we describe in detail our implementation and evaluate it using real-world data, including a performance comparison with an emulator, showcasing the performance-tradeoffs that practitioners must face.

\textbf{Roadmap.} The rest of this paper is organized as follows.
Section~\ref{sec:background} provides relevant background material on TEEs, \tz and trusted applications in general, including some of the underlying libraries and systems used in our evaluation.
We present the architecture of \sys in Section~\ref{sec:architecture}, detailing some of its implementation details in Section~\ref{sec:implementation}.
Section~\ref{sec:evaluation} presents our in-depth performance evaluation of the \sys prototype, using micro- and macro-benchmarks as well as real-world data.
We survey related work in Section~\ref{sec:related-work}, before concluding and devising future work in Section~\ref{sec:conclusion}.

 
\section{Background}\label{sec:background}

\textbf{Trusted Execution Environments (TEEs)}. A trusted execution environment is a hardware-protected part of the processor.
Depending on the specific version and implementation, a TEE can guarantee
confidentiality, integrity and protection against several types of attacks for code and data executed and processed within it.
Currently, there exist several hardware-based technologies that enable physical isolation of different execution environments available in a wide range of CPUs, including \arm \tz~\cite{trustzone,pinto2019demystifying,amacher2019performance}, \sgx~\cite{costan2016intel}, AMD SEV~\cite{kaplan2016amd}, and RISC-V Keystone~\cite{lee2020keystone}.
While we expect more TEE options to surface in the coming years, we focus on \arm\tz in the reminder of the paper, highlighting its main features and programming framework.

\textbf{\tz}. \tz is a hardware feature implemented in \arm processors since 2004~\cite{alves2004tz}. It enables physical separation between two different execution environments: the trusted side (known as the TEE or \emph{secure world}) and the untrusted side (known as the REE or \emph{normal world}). The \tz protects the integrity and confidentiality of the code run inside the \emph{secure world} from an attacker with physical access to the device, a malicious kernel or a high-privileged software. Programs hosted inside the \tz, known as Trusted Applications, can leverage additional \tz functionalities such as secure persistent storage with the use of APIs.

\begin{figure}[!t]
  \centering
  \includegraphics[scale=0.7]{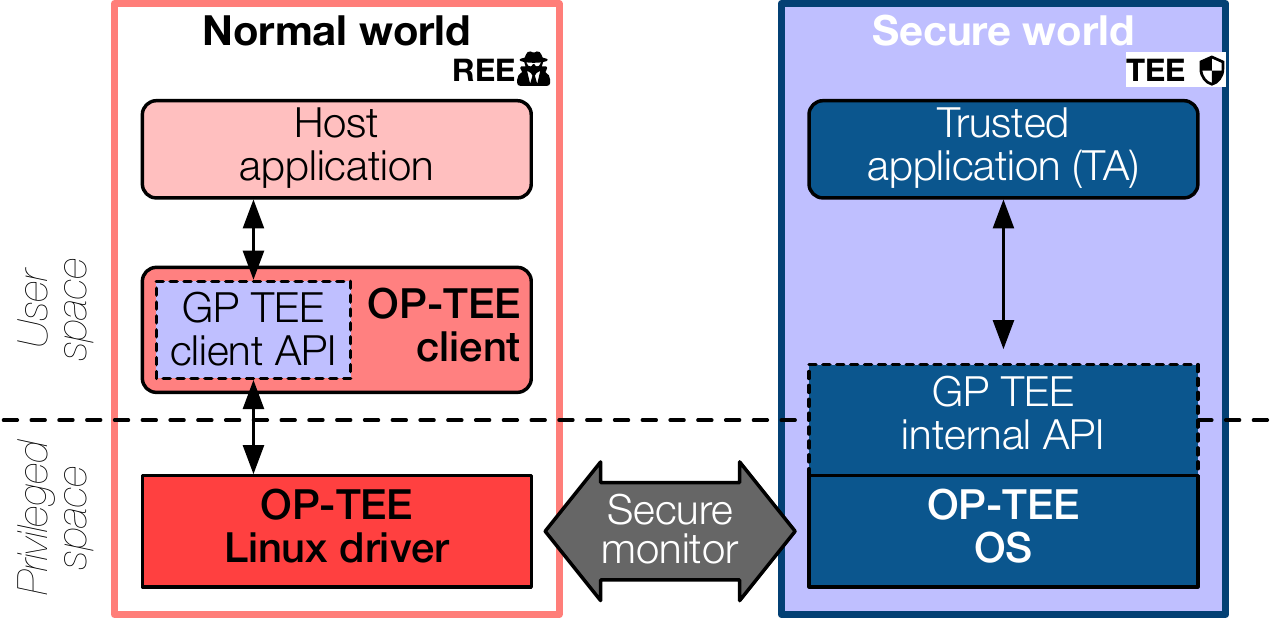}
  \caption{Architecture of \optee to realize trusted applications leveraging \tz. GP: GlobalPlatform~\cite{globalplatform}, a standardization effort for TEEs.}
  \label{fig:optee}
\end{figure}

\textbf{\optee}. The Open Portable Trusted Execution Environment (\optee) is an open source operating system with native support for the \tz. \optee implements two APIs compliant with the GlobalPlatform~\cite{globalplatform} specifications: the TEE Internal Core API~\cite{teeinternalapi}, which is exposed to the Trusted Applications, and the TEE Client API~\cite{teeclientapi}, which defines how a client in the REE should communicate with the TEE. The TEE can run alongside a Linux-based operating system (such as a GNU/Linux distribution or AOSP) as the untrusted OS.

\textbf{Trusted Application}. Trusted Applications (TAs) run inside the \emph{secure world}, making use of the TEE kernel to access system resources. TAs can act as a service for applications running on the \emph{normal world} as well as for other TAs. When using \optee, Trusted Applications are implemented in C and they can leverage the TEE Internal Core API implemented by \optee, which offers several services, including trusted storage and cryptographic, time and arithmetical operations. \sys is an application that runs on the TEE, so it is a Trusted Application. When using \sys, we can do so from another TA (if we are running it on the TEE) or from a normal application running in the REE. The design of trusted applications for \optee is depicted in Figure~\ref{fig:optee}.

\textbf{Trusted Persistent Storage}. \optee provides the Trusted Storage API for Data and Keys as part of the TEE Internal Core API~\cite{teeinternalapi}. This API can be used by Trusted Applications to access a secure storage which is only accessible to that particular TA and that is persistent between reboots. The data is stored encrypted and signed on the disk, to prevent it from being accessed or tampered with by any other application. The data can later be transparently accessed in cleartext by the TA. \sys exploits this by saving the encryption keys using a public ID, which are the value and key (respectively) in the key-value storage. When an untrusted application needs to use an encryption key, it sends the key's ID and \sys retrieves it.

\textbf{\mqtt \& \mosquitto}. The Message Queuing Telemetry Transport (\mqtt) is a lightweight, publish-subscribe network protocol, suited for communication in environments with few resources and low network bandwidth. MQTT has two types of entities: the broker and the clients. The clients can publish messages to a topic or subscribe to one of them, while the broker is a server that forwards each incoming message to all the subscribers of its topic. \mosquitto is an open source implementation of the \mqtt broker developed and maintained by the Eclipse Foundation, which also provides a C library for implementing MQTT clients, as well as one implementation of both a subscriber client and a publisher client.

\textbf{\mqttz}. \mqttz \cite{segarra2020mqttz-srds} is a fork of \mosquitto, a topic-based publish-subscribe framework for IoT. It allows brokers and the clients to leverage the \tz TEE, by encrypting the messages sent on the network to prevent the broker from being able to read them, while maintaining the publish-subscribe pattern. Similarly it allows full decoupling between publishers and subscribers, shielding the subscriptions inside the TEE.

\begin{figure}[t]
  \centering
  \includegraphics[scale=0.7]{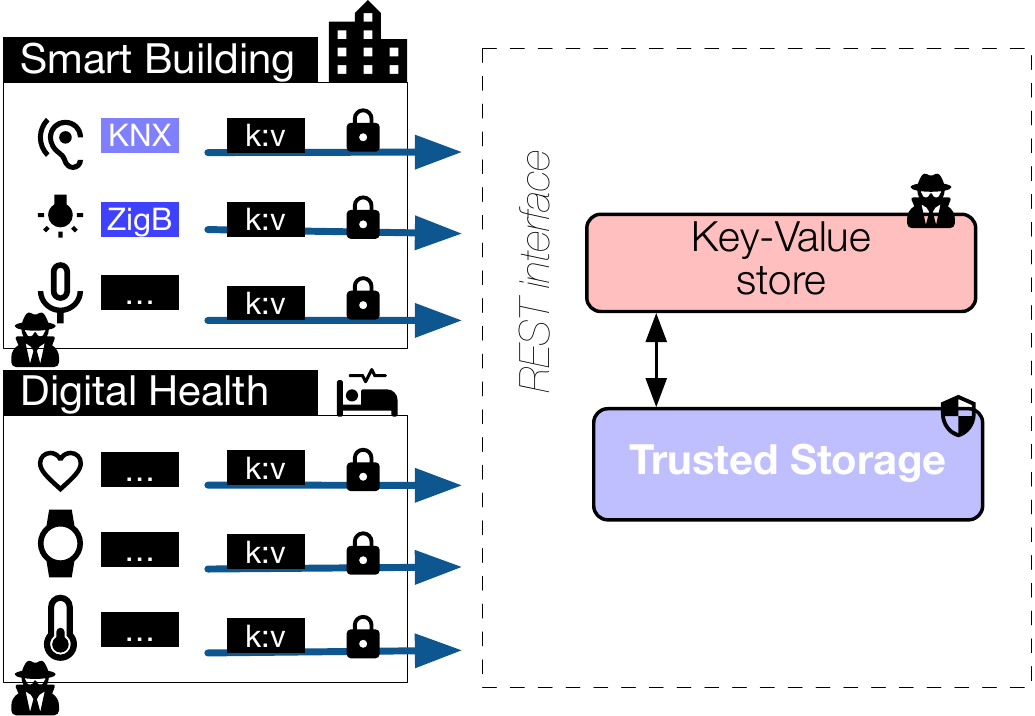}
  \caption{Two possible application scenarios where a trusted key/value storage system is valuable. Clients (on the left-side) issue requests to store key-value pairs into the key-value store, which stores those into a trusted storage.}
  \label{fig:kevlar:scenarios}
\end{figure}
\section{Motivating Scenarios} \label{sec:scenarios}

In this section, we describe our two main real-world scenarios behind \sys, also depicted in Figure~\ref{fig:kevlar:scenarios}.
The use-cases originate from two ongoing EU H2020 projects, in collaboration with industry-leading companies, which we details next.

\subsection{Digital Health}\label{ssec:health}

The first scenario stems from the H2020 project MOORE4MEDICAL.\footnote{\url{https://moore4medical.eu/}}
One of its objectives is to use wearable sensors and remote sensing technologies to reduce hospitalization, resulting in more comfort for the patient and less costly clinical trials in drug development.
In this context, the monitoring of vital signs is increasingly off-loaded and out-sourced to third-party untrusted data centers.
The main reason for such off-load is to exploit the economy of scale that comes with cloud computing.
The flow of data is mainly generated from smart medical devices and sensors and it is composed of a mix of physiological signals (\eg, electrocardiograms, photopletymograms) and vital signs (\eg heart rare, respiration rate, stress levels).
The data streams are highly heterogeneous, since the physiological signals can reach high sampling rates (\eg, Holter operate at 1~kHz) and the vital signs have typically much lower sampling rates ($\sim$1~Hz).

\subsection{Smart Building Management}

The second scenario stems from TABEDE\footnote{\url{http://www.tabede.eu/}}, an EU H2020 project with the aim to integrate energy grid demand-response schemes into buildings through low-cost extenders for Building Management Systems or as a standalone system, which is independent from communication standards and integrates innovative flexibility algorithms.
The flow of information relies on MQTT brokers deployed at the edge to minimize latency and limits the physical access from untrusted entities.
However, it directly raises several privacy and security concerns.
The flow of data is mainly generated from home appliances and sensors and it is composed of physical magnitudes such as electric current, temperature, or humidity. The data streams are generated at a slow frequency ($<$1~Hz) and are transfered via a large variety of communication protocols (\eg, EnOcean~\cite{li2014bacnet}, KNX~\cite{lee2009implementation}, Zigbee~\cite{farahani2011zigbee}).

\vspace{-10pt}
\section{Architecture} \label{sec:architecture}
\begin{figure}[!t]
  \centering
  \includegraphics[scale=0.7]{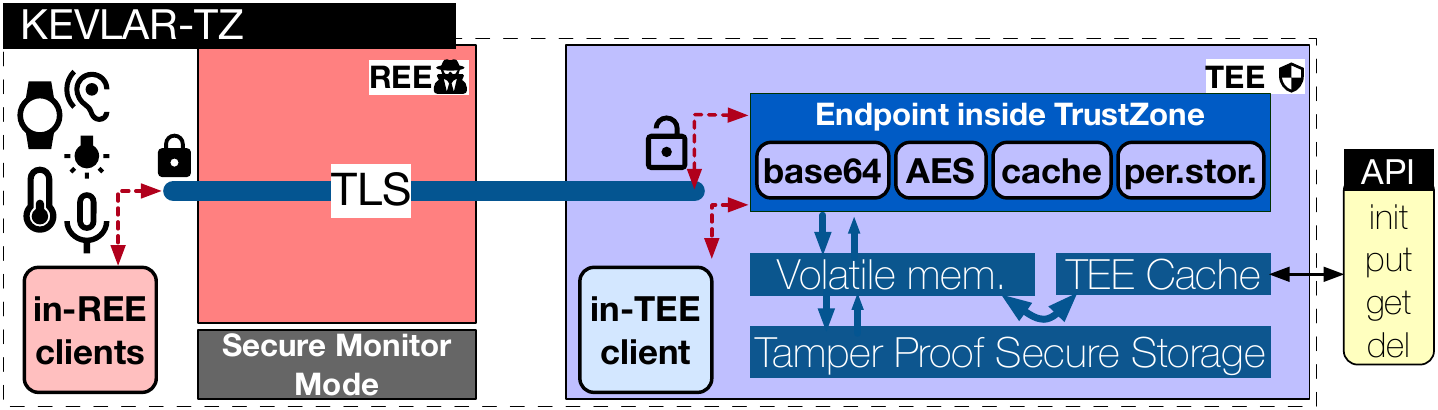}
  \caption{Architecture of the \sys TA.}
  \label{fig:kevlar:arch}
\end{figure}

\sys implements a secure key-value storage with non-volatile entries (\ie, available across reboots). 
To do so, we leverage \optee's Trusted Storage API~\cite{optee-securestorage} to store keys to a secure persistent storage, while implementing a cache in volatile memory to minimize the number of requests made to the persistent storage. 
The key idea is to limit as much as possible operations (\ie, \texttt{read}/\texttt{write}) involving the persistent storage, as they are considerably slower (see our evaluation in Section~\ref{sec:evaluation}). 

The architecture of \sys is depicted in Figure~\ref{fig:kevlar:arch}. 
In the remainder of this section, we describe the designing principles behind its various components as well as their interaction.
Finally, we detail the typical workflow of a single \texttt{write} operation, a keystone operation of \sys.

\textbf{Secure Persistent Storage}. The persistent storage area is a dedicated hardware component that guarantees data durability, confidentiality and integrity. 
\optee supports two modes for secure storage: \emph{(1)} using the REE file-system (the default option), or \emph{(2)} relying on a Replay Protected Memory Block (RPMB) partition of an eMMC device~\cite{reddy2015mobile}. \sys uses the REE file-system.

\sys implements a wrapper around the Trusted Storage API to access directly to writing and reading operations, which otherwise requires the management of several internal \optee components (omitted from the Figure~\ref{fig:kevlar:arch} for the sake of clarity).

This wrapper exposes two functions:
\begin{itemize}
    \item \texttt{read\_ss(const char *key, char *value, uint32\_t *value\_sz)}: reads the data mapped to a given \texttt{key}, which is bound to the array pointed by \texttt{value};
    \item \texttt{write\_ss(const char *key, const char *value, uint32\_t value\_sz)}: writes the data in \texttt{value} mapped to the \texttt{key} into persistent storage.
\end{itemize}

Finally, we note that the available storage memory dedicated to this component is only limited by the underlying hardware.

\textbf{Volatile memory -- Cache}. This component is the secure caching component of \sys.
Our design supports a few cache eviction policies (currently limited to Least Recently Used LRU and  FIFO). 
The implementation uses structs inserted in a queue and a hash table.
These are used to handle the key and value of each entry. 
The queue is used to remember the order of deletion of entries when new entries are to be added to a full cache; the hash table is used to access entries in average constant time. 
The cache is \texttt{write-through}~\cite{jouppi1993cache}, so that if the trusted application is stopped unexpectedly, no data is lost.

\textbf{API for Trusted Applications}. \sys provides a very simple API for applications running inside the TEE with four operations:

\begin{itemize}[noitemsep,topsep=0pt,parsep=0pt,partopsep=0pt]
    \item Initialize a cache with a given configuration consisting of cache size, hash output size and eviction policy;
    \item Delete a cache, freeing all space used in volatile memory. Objects in persistent storage are left untouched.
    \item Query a cache, to fetch the value associated to a given key. For instance, when using \mqttz~\cite{segarra2020mqttz-srds}, for a given ID, the cache will return the corresponding encryption key.
    \item Save a new key/value pair in volatile and persistent memory. 
\end{itemize}

\textbf{TCP interface for applications of REE}. The TEE and REE are two different systems and, as such, programs can't communicate (\ie, share data) between each other as if they where running on the same machine. 
However, \sys can be useful as a secure cache service to an application running in the \emph{normal world} (\eg, in the \mqttz broker scenario~\cite{segarra2020mqttz-srds}).
To expose \sys to the \emph{normal world}, we designed and implemented a TCP interface, protected by \tz, that allows to communicate \sys with any other application reachable on the network.

The establishment of the TCP connection works as follows.
First, an application in the REE opens a server TCP socket.
Secondly, \sys connects to such socket and waits (\ie, blocks) for new messages. 
Once a new message is received, \sys will execute the requested operation and return the desired value.

\textbf{The workflow of a write}. 
To conclude the description of the architecture, we take a step-by-step walkthrough for a \texttt{write} operation to insert a new key/value pair into \sys.
When an REE-based application needs to store a new key/value, it must first connect to \sys via TCP, and pass over the content of the key/value pair.
For the sake of simplicity, we assume those to be encoded using \texttt{base64}. 
Once received by the \sys TA, they get base64-decoded, and saved to the persistent storage.
The architecture allows to attach additional application-specific processing operations to the inserted key/value pairs, both before or after the value is retrieved.
For instance, one might send a cipher that will be decrypted with one value and encrypted with another~\cite{segarra2020mqttz-srds}, securely changing the encryption key of a cipher without the REE ever getting ahold of any of them.
This post-retrieve operations can be changed to any operation needed for the application that is using \sys.

If an application in the \emph{secure world} uses \sys to store a new key, it can directly leverage the functions exposed by the \sys API  (\texttt{cache\_save\_object(Cache *cache, char *id, char *data)}), which takes the binary values and stores them to the persistent storage.



\section{Implementation}\label{sec:implementation}
This section describes some of the internal details and implementation \sloppy{choices of \sys}.
The system itself is implemented in C, and consists of 791 LoC, released as open-source from \url{https://github.com/mqttz/kevlar-tz/}.
We note that applications implemented using the \optee framework are basically organized as two distinct components: the Host Application (HA) and the corresponding Trusted Application (TA). 
The host application runs on the \emph{normal world} and initilizes and finalizes the TEE context using the TEE Client API. 
Moreover, the HA is in charge of invoking functions over the Trusted Application, and can do so multiple times, dividing the work between the \emph{normal} and \emph{secure world}. 
However, the TEE's volatile memory is lost between calls, hence \sys's Host Application only invokes the TA once.
The TA acts as a daemon that receives queries.
We detail the TA components next.

\subsection{\sys Trusted Application}
The \sys trusted application is split into several modules.
The implementation of a particular module is independent of the rest, to facilitate future evolutions of the code in a loosely coupled manner (\eg you can change the communication module to work with UDP instead of TCP, different symmetric encryption algorithm, \etc). 
The \sys TA is composed of the following modules, which we evaluate individually and as a part of micro-benchmarks in Section~\ref{sec:evaluation}.

\textbf{Persistent storage module}. This module implements the functions \texttt{read\_ss} and \texttt{write\_ss}, which read and write persistent storage, respectively. 
Our implementation follows the guidelines from the Linaro Security Working Group.~\footnote{\url{https://github.com/linaro-swg/optee_examples/tree/master/secure_storage}}

\textbf{Cache module}. The \sys cache module directly implements the proper cache API, namely: \texttt{init\_cache}, \texttt{free\_cache}, \texttt{cache\_query}, and \texttt{cache\_save\_object}.
In our implementation, the cache is made of nodes that are part of both a queue and a hash table, which enables accessing objects in constant time (on average). The cache module also interacts with the persistent storage (using the pertinent module).
Any access to the key-value storage can be done through the cache, whether the value was stored on volatile memory or not.

\textbf{AES module}. Our prototype includes a symmetric cipher module on top of AES. 
It directly exposes two functions: \texttt{encrypt}, which encrypts data with a given key, and \texttt{reencrypt}, which given two keys and a cipher, decrypts it with one key and encrypts it with the other. 
Our implementation uses the Cryptographic Operation API implemented by \optee to encrypt and decrypt the data.
We extended it to support PCKS padding, \ie the default padding used by OpenSSL and \mqttz.

\textbf{Base64 module}. This module implements standard Base64 encoding/decoding operations (\ie, \texttt{base64\_encode}, \texttt{base64\_decode}), as well as auxilary ones (\ie, \texttt{base64\_decode\_length} to return the length an encoded string after the decoding). 
The encoding and decoding implementations leverage an open-source library.\footnote{\url{https://web.mit.edu/freebsd/head/contrib/wpa/src/utils/base64.c}} 
This module is used to encode and decode data transmitted to other applications over the network layer to simplify the parsing of data, in particular when dealing with multipart binary messages.

\textbf{Trusted TCP module}. \sys uses TCP to communicate with untrusted applications. 
It exposes functions to initialize and close a connection (\ie, \texttt{net\_connect} and \texttt{net\_disconnect}) as well as send and receive packets (\ie, \texttt{net\_send} and \texttt{net\_receive}). 
The implementation uses the socket library that \optee exposes, and while we use TCP, the code can easily be adapted to use other protocols without any changes on any other part of the application.


\section{Evaluation} \label{sec:evaluation}





This section presents our experimental evaluation of \sys using both micro- and macro-benchmarks.
Our goal is to define the overheads of running \sys to further assess whether the trade-offs to have a secure storage system are reasonable for a real-world scenario.

\textbf{Evaluation settings}. We deploy \sys on a Raspberry Pi 3 Model B+ as well as on an emulated environment using QEMU version 8~\footnote{\url{https://www.qemu.org}} to test the application.
QEMU is a tool that has been proven useful, despite its limitations, in validating design and implementation in \arm processors without having to deploy large (and potentially) expensive testbeds.
The QEMU runtime is deployed on a Lenovo ThinkPad with Intel\textregistered{} Core\texttrademark{} i7-5600U CPU @ 2.60GHz.
We rely on OP-TEE version 3.11.0.

\subsection{Micro-benchmarks}
We begin by micro-benchmarking two of the subcomponents of the \sys TA, namely the Base64 encoder and the one in charge of cryptographic operations. 

\begin{figure}[!t]
    \centering
    \includegraphics[scale=0.7]{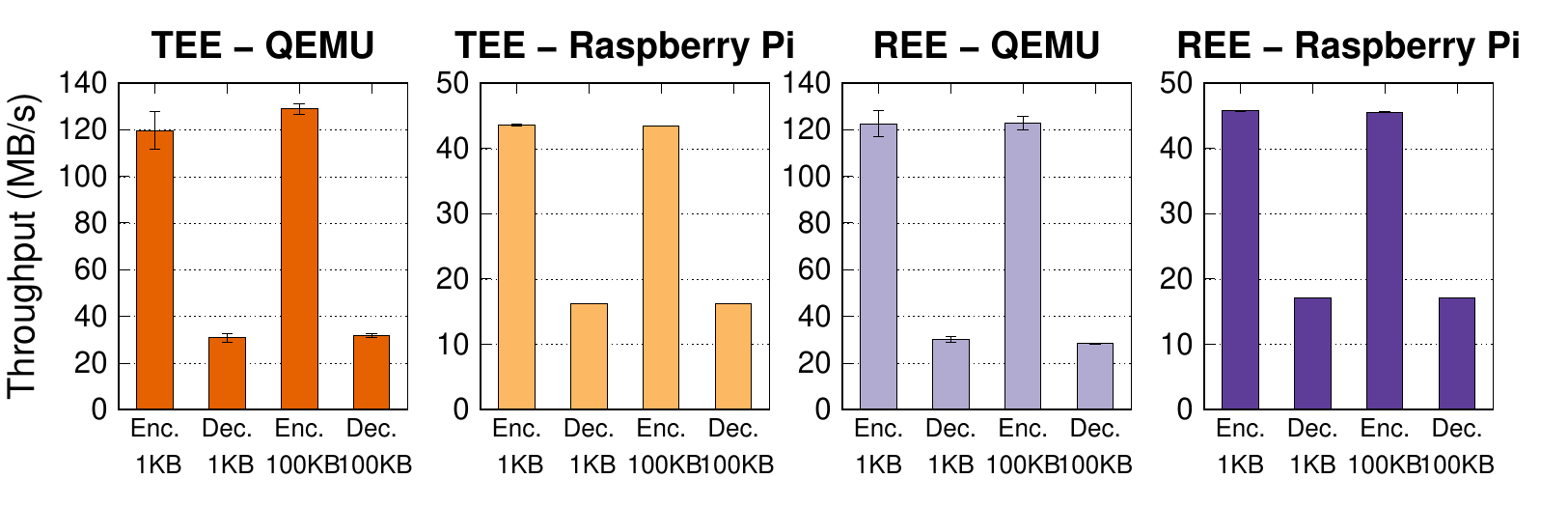}
    \caption{Base64 encoding and decoding throughput for randomly generated data.}
    \label{fig:base64}
\end{figure}

\textbf{Base64 encoding and decoding}. We measure the throughput of the base64 encoding and decoding operations. 
The measurements have been done by measuring the encoding and decoding of randomly generated data of 1KB and 100KB, both for the hardware deployment as well as under emulation.
In both cases, the component is deployed in the TEE.
We show average and standard deviation for each configuration, which is executed 200 times.
Our results are shown in Figure~\ref{fig:base64}.
First, we observe the size of the data payload does not negatively affect the observed throughput, whereas we do observe differences between the emulated and hardware environment.
For instance, encoding 100KB of data reaches 43 MB/s, while the QEMU is approximately 4$\times$ faster, reaching 130MB/s.
Similar differences can be observed for smaller data and decoding.
We also report the results obtained when executing the same operations in the REE. As we see, encoding and decoding throughputs are similar, due to the operations being GPU-intense, instead of memory-intense.

\begin{figure}[!t]
    \centering
    \includegraphics[scale=0.7]{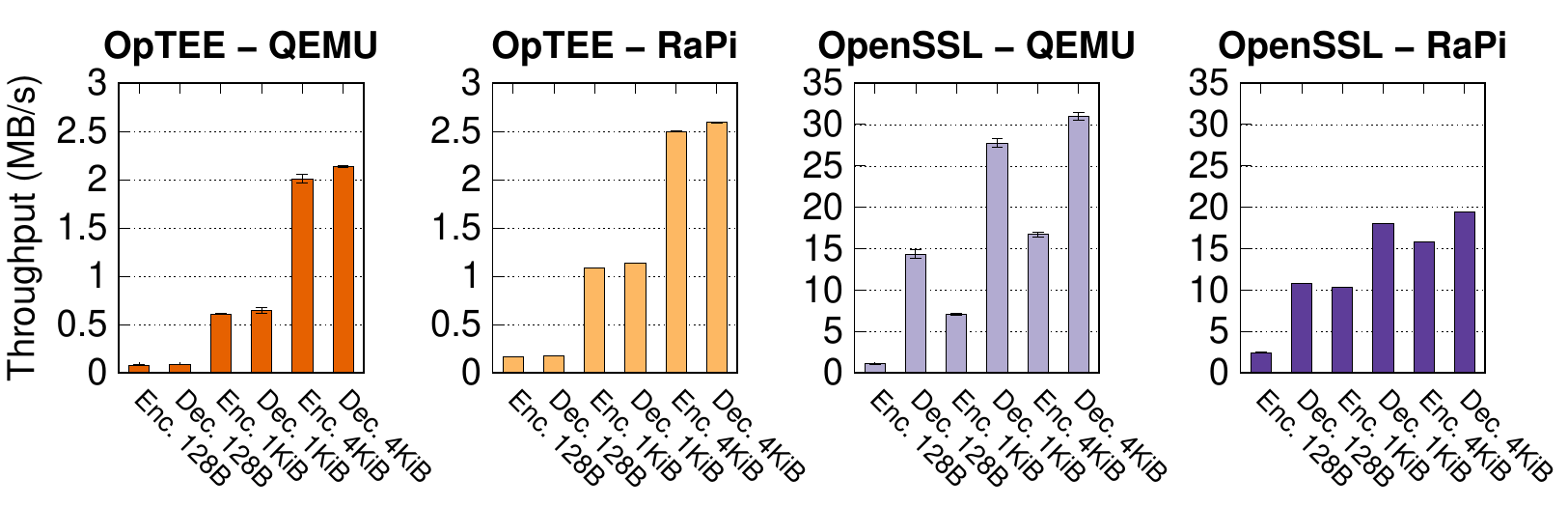}
    \caption{Encryption and decryption throughput. We compare the built-in \optee library for cryptographic operations against OpenSSL. For both cases we deploy them in TEE, and compare the throughput to encode and decode randomly generated data.}
    \label{fig:crypto}
\end{figure}

\textbf{Cryptographic operations}. Next, we measure the throughput of the cryptographic operations run by \sys, \ie symmetric encryption and decryption. 
We generate random data of different sizes: 128B, 1kB and 4kB. 
Figure~\ref{fig:crypto} depicts our results.
We observe that encryption and decryption achieve similar encoding and decoding throughputs in each of the two environments (QEMU and the Raspberry Pi).
For instance, we observe an average of 2 MB/s encrypting a payload of 4kB in QEMU, and a 25\% improvement for the same test in hardware.
Expectedly, decryption operations are slightly faster (by 6\% on average).
We compare our results with OpenSSL version 1.1.1f, running on the REE of both QEMU and Raspberry Pi.
We observe that OpenSSL in the REE is much faster, especially for decryption operations which can be parallelized: this is expected, as OpenSSL optimizes the compiled binary for the underlying hardware.
We leave as future work further investigation and porting of a (subset of) OpenSSL to run in the TEE.
 

\textbf{TCP communication}. The TPC sockets handle the communication between \sys and untrusted applications.
In this benchmark, we measure the throughput of our trusted TCP channels, whose endpoints terminate into the \tz area.
We measure the throughput for messages of different sizes: 1B, 245B (\ie, 128B once encrypted and encoded in base64), and  757B (\ie, 512B plus AES encryption and base64 encoding), and 1024B.
We use these values since they represent a reasonable range of values found in real-world deployments.
Figure~\ref{fig:tcp} reports our results for the two testing environments. 
We observe that the throughput is significantly higher for larger amounts of data. Concerning the system used, we see that the Raspberry Pi is much faster than the emulated environment.

\begin{figure}[!t]
    \centering
    \includegraphics[scale=0.7]{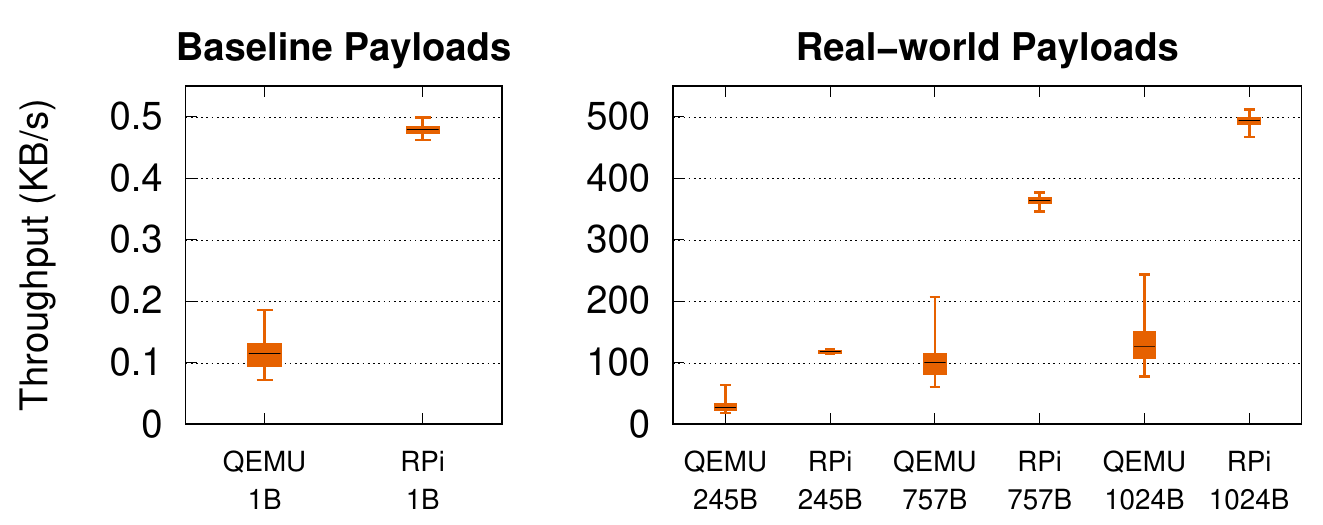}
    \caption{Trusted TCP server: incoming throughput.}
    \label{fig:tcp}
\end{figure}

\begin{figure}[!t]
    \centering
    \includegraphics[scale=0.7]{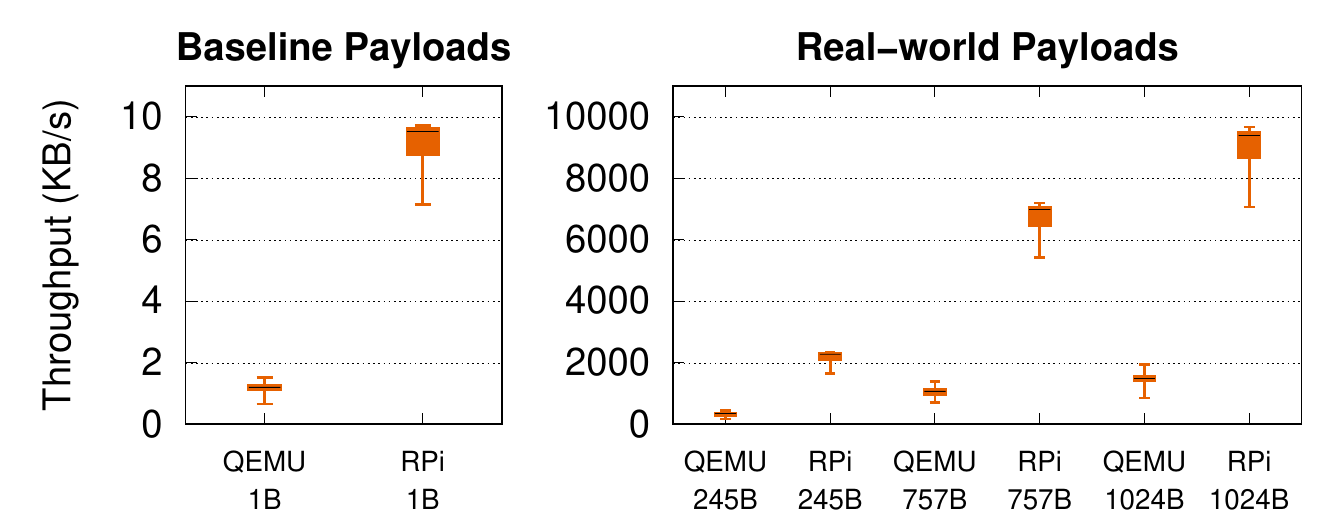}
    \caption{TCP throughput REE to REE \ob{prettify plot and caption}}
    \label{fig:tcp}
\end{figure}

\begin{figure}[!t]
    \centering
    \includegraphics[scale=0.7]{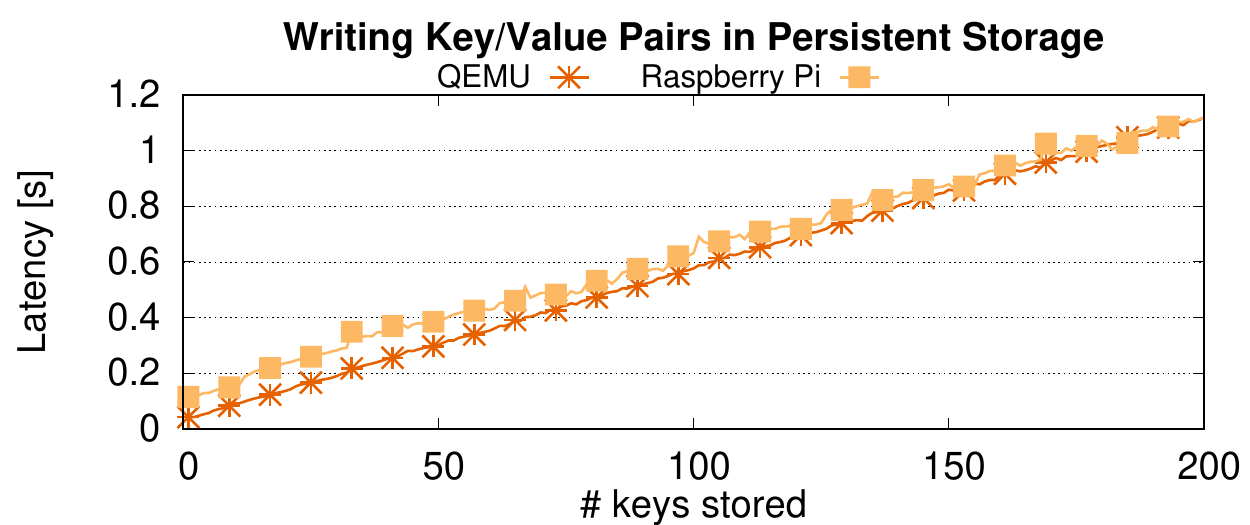}
    \caption{Time to store a 32B encryption key (value) with a 12B ID (key) to persistent storage.}
    \label{fig:persistent-store}
\end{figure}

\textbf{Cache module}. The last in our series of micro-benchmarks focus on the throughput of the cache module itself.
First, we fill up the persistent storage with 200 keys, using a payload of 32B.
Figure~\ref{fig:persistent-store} indicates that as the number of keys increase, there is a corresponding increase in the time to insert the key and value in the persistent storage.

\begin{figure}[!t]
    \centering
    \includegraphics[scale=0.7]{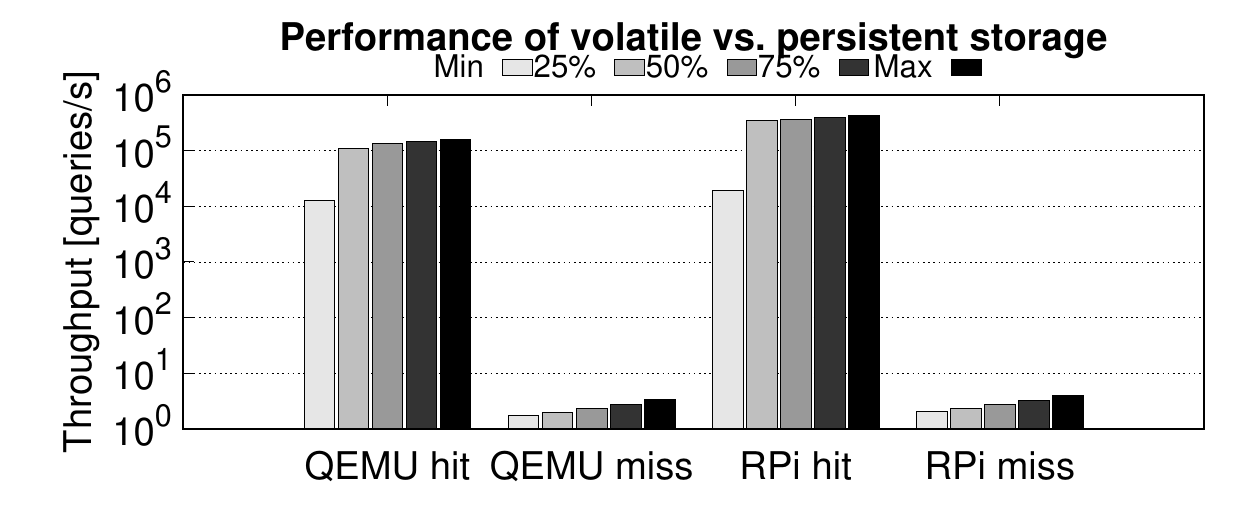}
        \caption{Throughput of queries. We compare the difference between volatile memory (cache hits) and persistent storage (cache miss) on both hardware (Raspberry Pi) and emulated (QEMU) environments. The percentiles of the distribution are represented with shades of gray. From the brightest to the darkest: minimum, 25th, the 50th (median), the 75th and the maximum percentile.}
    \label{fig:cache-get}
\end{figure}

Finally, we query the previously stored keys by issuing request queries for random keys to the cache module.
Each cache request can result in a \emph{hit} or \emph{miss}. 
These results are shown in Figure~\ref{fig:cache-get}.
We can observe 5 orders of magnitude between the hits and miss throughputs, both in the case of emulated as well as hardware deployments. 


\subsection{Macro-benchmark: Digital Health}
\begin{figure}[t]
    \centering
    \includegraphics[width=0.7\linewidth]{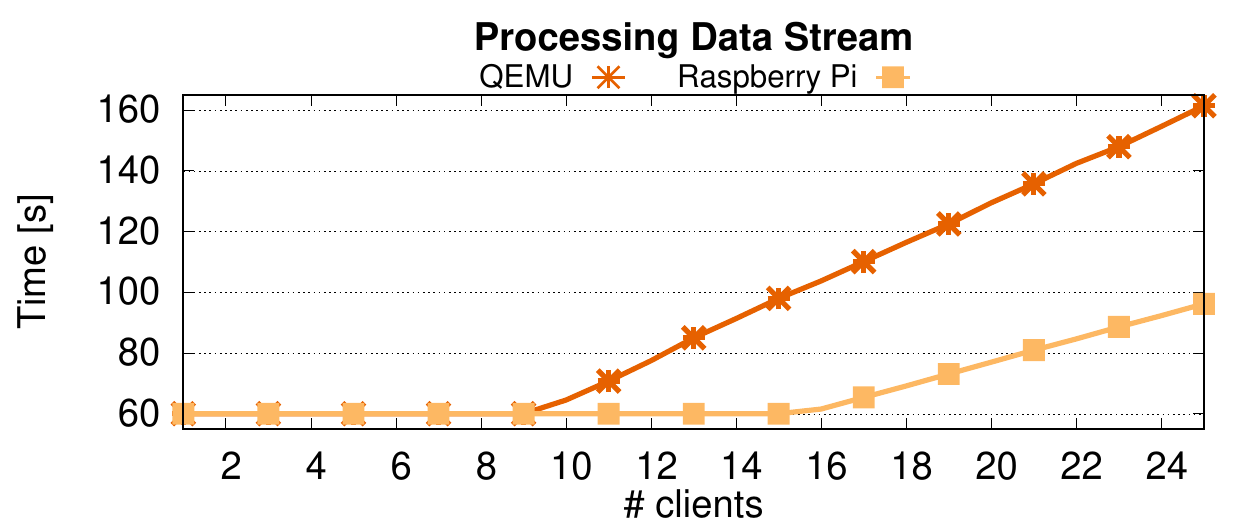}
    \includegraphics[width=0.29\linewidth]{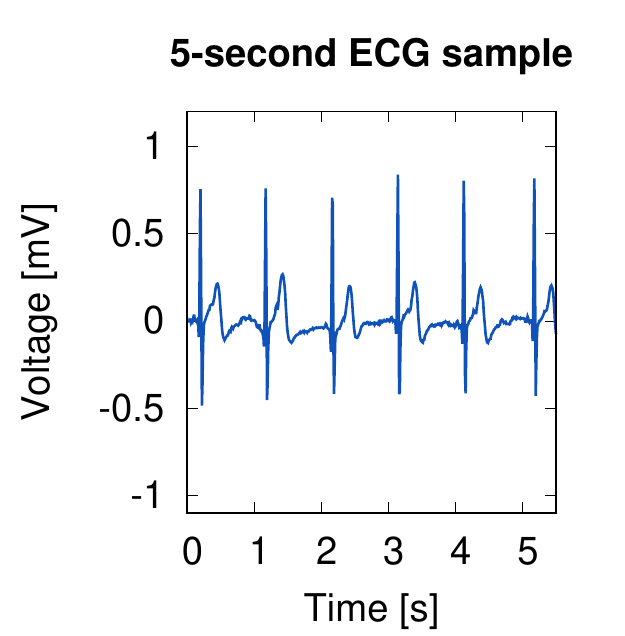}
    \caption{Time to process a 60-second data stream from ECG sensors.}
    \label{fig:macro}
\end{figure}

We conclude our evaluation by demonstrating the overall performance of \sys. 
We setup the Digital Health scenario (Section~\ref{ssec:health}), where heart-rate monitoring data streams are pushed toward \sys, leaving the Smart Building use-case to future work.
For the considered workload, we use a database obtained from CSEM’s proprietary wrist-located sensors and chest-located dry electrodes~\cite{chetelat2015}.
In particular, cardiac data is obtained following a standardized protocol in which they perform a range of physical activities from sedentary to vigorous~\cite{delgado2014}.
A 5-second sample of the ECG used is shown on Figure~\ref{fig:macro}.
In this scenario, the sensors inject 10 electrocardiogram data points every 93.4ms.
Each data point embeds the following information: the time when it was taken and the voltage measured.
Plot~\ref{fig:macro} compares the time taken to process 60 seconds of such streaming for different amounts of clients. 
\vs{maybe this plot has to be changed a bit, to be discussed tomorrow. Since this is a 'scalability' plot, we should show plot a bit differently the data. On the y-axis you can put the average time to process 1 second of message. Maybe it is sufficient to divide /60 the y-axis values.}
We conclude that the Raspberry Pi takes approximately 0.064 seconds to process 1 second of streaming of 1 client, while the emulated environment takes 0.107 seconds for the same amount of data.

\vspace{-10pt}

\section{Related Work}\label{sec:related-work}
The problem of executing software inside TEEs in general, and \tz in particular, has attracted several research groups.
CaSE~\cite{zhang2016case} is a cache-assisted secure execution framework for the \tz to defend against multiple attacks.
Others~\cite{havet2017securestreams,park2019streamboxtz,segarra2019meddata} have implemented frameworks for TEEs to securely process data streams that could benefit from \sys.
However, while such projects have implemented full-fledged frameworks, \sys provides a lean and resource-efficient cache with an easy-to-use API for applications that need fast access to persistent data.


A service like \sys is available within \mqttz~\cite{segarra2020mqttz-srds}, a publish-subscribe framework optimized for IoT and \tz deployments and backward compatible with \mosquitto, a well-known MQTT messaging framework supporting TLS.
In that context, a secure cache like the one developed in \sys protects data against eavesdroppers or untrusted brokers.
\sys offers a more generic approach, including an API to use it from inside the TEE and a modular design to choose a specific cache eviction policy or some of its internal subcomponents.
Recently~\cite{sasaki2020secure,wan2020rustee}, authors tried to hardening TEE applications against a broad set of attacks, including side-channels or against known weaknesses of the implementation language.
While some of the countermeasures developed there could be beneficial for \sys, we consider those out of scope.
In~\cite{gentilal2017bitcoin}, authors implement a cache to speed up operations on a secure Bitcoin wallet, while using the \tz's persistent storage.
While the focus is on the security of the private keys used to unlock the cryptocurrency wallet, the approach is similar to \sys.

To the best of our knowledge, \sys is the first application specifically designed to run on a \tz and provide a lightweight cache to leverage the \tz's persistent storage while maintaining a minimal read/write latency.
\sys implements a generic cache that can be easily embedded into other Trusted Applications or used as a secure storage for an untrusted application in the \emph{normal world} without significantly increasing the trusted computing base.

\vspace{-10pt}
\section{Conclusion and Future Work} \label{sec:conclusion}
\vspace{-6pt}

Motivated by the increasing attack surface of today's edge devices, \sys addresses an integral part for storing and performing computations over the collected/transmitted data in a privacy-preserving fashion. This becomes critical when the data is highly sensitive and personal, which is the case for nowadays medical implantables, wearables, and nearables. 
For instance, in scenarios where such IoT devices interact by means of publish/subscribe frameworks, as is typical in real-world deployments, protecting the brokers with a minimal increase in power consumption is necessary in order to preserve the ubiquity of such network of sensing devices.

We intend to extend \sys along the following lines.
First, we intend to extend the API exposed to the Trusted Application so that it is easier to implement new functions (similar to the already implemented reencryption.
Second, standardizing the length of the messages going through TCP when communicating with untrusted applications, so that binary data can be sent and easily parsed, this will reduce the amount of bytes sent as well as eliminate the base64 dependency.
Finally, we intend to compare \sys to other \tz cache implementations.

\vspace{-10pt}

\section{Acknowledgements}\label{sec:acks}

This work is supported in part by Moore4Medical, which has received funding within the Electronic Components and Systems for European Leadership Joint Undertaking (ECSEL JU) in collaboration with the European Union's H2020 framework Programme (H2020/2014-2020) and National Authorities, under grant agreement H2020-ECSEL-2019-IA-876190. Moreover, this project has received funding from the European Union's Horizon 2020 research and innovation programme under grant agreement No 766733.


     \bibliographystyle{splncs04}
     \bibliography{biblio}

\end{document}